\documentclass[12pt]{article}
\usepackage{epsfig}

\makeatletter
\renewcommand\section{\@startsection {section}{1}{\z@}%
                                   {-3.5ex \@plus -1ex \@minus -.2ex}%nn
                                   {2.3ex \@plus.2ex}%
                                   {\normalfont\large\bfseries}}
\renewcommand\subsection{\@startsection{subsection}{2}{\z@}%
                                     {-3.25ex\@plus -1ex \@minus -.2ex}%
                                     {1.5ex \@plus .2ex}%
                                     {\normalfont\bfseries}}
\makeatother
\def\IZ{\relax\ifmmode\mathchoice
{\hbox{\cmss Z\kern-.4em Z}}{\hbox{\cmss Z\kern-.4em Z}}
{\lower.9pt\hbox{\cmsss Z\kern-.4em Z}} {\lower1.2pt\hbox{\cmsss
Z\kern-.4em Z}}\else{\cmss Z\kern-.4em Z}\fi}
\def\IR{\relax{\rm I\kern-.18em R}}

\def\one{{\hbox{ 1\kern-.8mm l}}}

\newlength{\bredde}
\def\slash#1{\settowidth{\bredde}{$#1$}\ifmmode\,\raisebox{.15ex}{/}
\hspace*{-\bredde} #1\else$\,\raisebox{.15ex}{/}\hspace*{-\bredde}
#1$\fi}

\newsavebox{\zzzbar}
\sbox{\zzzbar}
  {\setlength{\unitlength}{0.9em}
  \begin{picture}(0.6,0.7)
  \thinlines
  \put(0,0){\line(1,0){0.6}}
  \put(0,0.75){\line(1,0){0.575}}
  \multiput(0,0)(0.0125,0.025){30}{\rule{0.3pt}{0.3pt}}
  \multiput(0.2,0)(0.0125,0.025){30}{\rule{0.3pt}{0.3pt}}
  \put(0,0.75){\line(0,-1){0.15}}
  \put(0.015,0.75){\line(0,-1){0.1}}
  \put(0.03,0.75){\line(0,-1){0.075}}
  \put(0.045,0.75){\line(0,-1){0.05}}
  \put(0.05,0.75){\line(0,-1){0.025}}
  \put(0.6,0){\line(0,1){0.15}}
  \put(0.585,0){\line(0,1){0.1}}
  \put(0.57,0){\line(0,1){0.075}}
  \put(0.555,0){\line(0,1){0.05}}
  \put(0.55,0){\line(0,1){0.025}}
  \end{picture}}

\newcommand{\ena}{\end{eqnarray}}
\newcommand{\beqa}{\begin{eqnarray}}
\newcommand{\eeqa}{\end{eqnarray}}
\newcommand{\bea}{\begin{eqnarray}}
\newcommand{\eea}{\end{eqnarray}}

\newcommand{\be}{\begin{equation}}
\newcommand{\ee}{\end{equation}}

\def\H{{\cal H}}
\usepackage{graphicx}

\newcommand{\beq}{\begin{equation}}
\newcommand{\eeq}{\end{equation}}
\newcommand{\ber}{\begin{array}}
\newcommand{\eer}{\end{array}}

\newcommand{\dsty}{\displaystyle}

\newcommand{\eps}{\varepsilon}

\usepackage{amsmath}
\usepackage{amssymb}

\begin{document}
\begin{titlepage}
\begin{flushright}
arXiv:0812.2900
\end{flushright}
\vfill
\begin{center}
{\LARGE\bf Can free strings propagate\vspace{2mm}\\
across plane wave singularities?}    \\
\vskip 10mm
{\large Ben Craps,$^a$ Frederik De Roo$^{a,b,}$\footnote{Aspirant FWO} and Oleg Evnin$^a$}
\vskip 7mm
{\em $^a$ Theoretische Natuurkunde, Vrije Universiteit Brussel and\\
The International Solvay Institutes\\ Pleinlaan 2, B-1050 Brussels, Belgium}
\vskip 3mm
{\em $^b$ Universiteit Gent, IR08\\Sint-Pietersnieuwstraat 41, B-9000 Ghent, Belgium}
\vskip 3mm
{\small\noindent  {\tt Ben.Craps@vub.ac.be, fderoo@tena4.vub.ac.be, eoe@tena4.vub.ac.be}}
\end{center}
\vfill

\begin{center}
{\bf ABSTRACT}\vspace{3mm}
\end{center}

We study free string propagation in families of plane wave geometries developing
strong scale-invariant singularities in certain limits. We relate the singular limit of the evolution for all excited string modes
to that of the center-of-mass motion (the latter
existing for discrete values of the overall plane wave profile normalization). Requiring that the entire excitation energy
of the string should be finite turns out to be quite restrictive and essentially
excludes consistent propagation across the singularity, unless dimensionful scales are introduced at the singular locus (in an otherwise scale-invariant space-time).

\vfill

%\begin{flushleft}
%PACS 11.25.-w, 04.65.+e
%\end{flushleft}
\end{titlepage}
%%%%%%%%%%%%%%%%%%%%%%%%%%%%%%%%%%%%%%%%%%%%%%%%%%%%%%%%%%%%%%%%%%%%%%%%%%%%%%%%%%%%%%%%%%%%%%%%%%%%%%%%%%%%%%%%%%%%%%%%%%%%%%%%%%%%%%%%%%%%%%%%%%%%%%%%%%%%%%%%%%%%%%%%%%%%%%%%%%%%%%%%%%%%%%%%%%%%%

\section{Introduction}

String propagation in strong gravitational waves has attracted a
considerable amount of attention on account of a few highly special
properties of such space-times (see \cite{HorowitzSteif,PRT} among other publications). For one thing, the structure of
the curvature tensor in plane gravitational waves implies that
these solutions to Einstein's equations (coupled to appropriate matter fields, if necessary) remain uncorrected \cite{HorowitzSteif,AmatiKlimcik}
in a number of higher derivative extensions of general relativity
(and, in particular, they do not receive any $\alpha'$-corrections
when introduced as backgrounds in perturbative string theories).
Furthermore, the corresponding light cone Hamiltonian of string 
$\sigma$-models turns out to be quadratic and admits a fairly
thorough analytic treatment. (This class of backgrounds
also admits a natural formulation of the matrix theory description
of quantum gravity \cite{MBB,ppmatrix}.)

In this publication, we shall concentrate on backgrounds
of the following form (so-called exact plane waves):
\be
ds^2=-2dX^+dX^- - F(X^+) \sum_{i=1}^d (X^i)^2 (dX^+)^2 + \sum_{i=1}^d(dX^i)^2.\label{genppwave}
\ee
This representation of the metric is often called the Brinkmann form.
The case of constant $F(X^+)$ corresponds to supersymmetric plane
waves studied in \cite{BMN}, and it is quite different from the
the rapidly varying $F(X^+)$ we intend to consider.
(A coordinate transformation can be performed into the so-called Rosen
coordinates eliminating the dependence of the metric on the transverse
coordinates $X^i$. The resulting metric depends on $X^+$ only and
displays manifestly a plane-fronted space-time wave propagating at the speed of light.
However, the Rosen parametrization tends to suffer from coordinate singularities,
and we shall work with the Brinkmann form.)

The function $F(X^+)$ contained in (\ref{genppwave}) is completely arbitrary,
and one may ask, for example, what happens to quantum strings propagating
in such space-times when the wave profile $F(X^+)$ develops an isolated
singularity. This question is of some interest per se, since
studies of string theory in the presence of space-time singularities
have played a pivotal role in the development of the subject (and, in this
particular case, we are dealing with singularities in time-dependent
backgrounds). Additional heuristic justification for our studies is
provided by the observation that plane waves of the type (\ref{genppwave})
with
\be
F(X^+)\sim \frac1{(X^+)^2}
\label{scaleinv}
\ee
arise as Penrose limits of a broad class \cite{plimit} of space-time singularities
(including the Friedmann-Lema\^\i tre-Robertson-Walker cosmological singularities).
With $F(X^+)$ of (\ref{scaleinv}), the metric (\ref{genppwave}) is
invariant under scaling transformations $X^+\to \alpha X^+$, $X^-\to X^-/\alpha$
(identical to Lorentz boosts in flat space-time). Note that this type of singularities is considerably stronger than the so-called ``weak singularities'' of \cite{David}.

Free string propagation on (\ref{genppwave}) with $F(X^+)$ given by (\ref{scaleinv})
has been previously studied in \cite{PRT}. In particular, it was suggested
in that publication that the question of propagation across the $1/(X^+)^2$
singularity in the metric can be addressed by employing analytic continuation
in the complex $X^+$-plane. We believe that this issue merits further elucidation.

In the context of string theory and related approaches to quantum gravity,
there is a general expectation that the space-time background used for
formulating the theory should satisfy some stringent consistency conditions.
For perturbative string theories, these conditions take the form of the
appropriate supergravity equations of motion together with an infinite
tower of $\alpha'$-corrections. For non-singular plane waves, all the
$\alpha'$-corrections vanish automatically on account of the special properties
of the Riemann tensor corresponding to these space-times. For singular space-times,
the question of background consistency conditions {\it at} the singular point
appears to be extremely subtle. Indeed, what should replace the supergravity
equations of motion at the singular point where they obviously break down?
Ad hoc prescriptions are not likely to produce meaningful results
under these circumstances.

One approach to formulating string theory in backgrounds (\ref{genppwave}-\ref{scaleinv}) is to resolve the singular plane wave profile
into a non-singular function, perform the necessary computations and
see if the result has a meaningful singular limit. (This approach was advocated in \cite{HorowitzSteif}, where a conjecture was made that for certain choices of the plane wave profile, taking a singular limit
may result in well-defined transition amplitudes. We intend to consider this question
quantitatively.) Note that, for the 
resolved space-times of this sort, perturbative string background consistency
conditions are automatically satisfied to all orders in $\alpha'$.
The only non-trivial question is the existence of a singular limit.

But how should one resolve? We want to construct a function $F(X^+,\epsilon)$
in such a way that
\be
\lim\limits_{\epsilon\to 0}F(X^+,\epsilon)=\frac{\mathrm{const}}{(X^+)^2}
\label{limt}
\ee
 everywhere away from $X^+=0$. There is in principle
a large amount of ambiguity associated with such resolutions. One class
appears to be very special however. The background (\ref{genppwave}-\ref{scaleinv})
possesses a scaling symmetry and does not depend on any dimensionful parameters.
It is natural to demand that this symmetry should be recovered when
the resolution is removed. This will happen if the resolved profile $F(X^+,\epsilon)$
does not depend on any dimensionful parameters other than the resolution
parameter $\epsilon$. In this case, on dimensional grounds,
\be
F(X^+,\epsilon)=\frac{\lambda}{\epsilon^2}\Omega(X^+/\epsilon).
\label{scaleinvres}
\ee
The limit (\ref{limt}) will be recovered if
\be
\Omega(\eta)\to \frac{k}{\eta^2}+O\left(\frac1{\eta^b}\right)
\label{asrefprof}
\ee
for large values of $\eta$, with some $b>2$.
Note that the fact that the original background possesses a certain symmetry
(away from $X^+=0$!) in no way implies that we {\it must} resolve in
a way consistent with this symmetry. For resolved profiles different from
({\ref{scaleinvres}), the limit of the metric may still be given by (\ref{limt})
away from $X^+=0$ (and thus be scale invariant), but additional dimensionful
scales may become buried inside the singularity at $X^+=0$ (in a way
that only affects processes involving singularity crossing).
One would need some strong physical rationale for introducing
such scales buried at the singular locus, and in the present publication
we shall simply study the ``scale-invariant'' resolutions (\ref{scaleinvres}).

The structure of the paper is as follows: we will first derive the Hamiltonian for a free string in the background (\ref{genppwave}-\ref{scaleinv}). Then we recapitulate the main results of \cite{EvninNguyen} for the evolution of the center-of-mass motion across the plane wave singularity. We extend this analysis to the evolution of excited string modes. We conclude by discussing stringent conditions arising if one 
demands the total mass of the string to remain finite after it crosses the
singularity.

%%%%%%%%%%%%%%%%%%%%%%%%%%%%%%%%%%%%%%%%%%%%5

\section{Free strings in plane waves}

Due to the presence of covariantly constant null vectors in plane wave geometries, the string theory $\sigma$-model
can be analyzed explicitly in such backgrounds, and reduces to a set of independent
classical time-dependent harmonic oscillators. In this section, we re-state this
familiar material in a way convenient for our present investigations.

\subsection{The light cone gauge}

String worldsheet fermions are free in plane wave backgrounds \cite{RT}. We shall therefore concentrate on the bosonic part of the string action, given by
\be
I=-\frac{1}{4\pi\alpha'}\int d\tau \int_0^{2\pi} d\sigma \sqrt{-g} \left(g^{ab} G_{\mu\nu} \partial_a X^\mu \partial_b X^\nu -\frac{1}{2} \alpha' R^{(2)}\Phi\right).
\label{bospartact}
\ee
We choose light-cone gauge $X^+=\alpha'p^+\tau$ and gauge-fix the metric,
\be
\mathrm{det}(g_{ab}) = -1,\hspace{5mm}\partial_\sigma g_{\sigma\sigma}=0,
\ee
to obtain the following Lagrangian, where we have solved for $g_{\tau\tau}$:
\begin{align}
L=-\frac{1}{4\pi\alpha'}\int_0^{2\pi}&\mathrm{d}\sigma \Biggl(2 g_{\sigma\sigma} p^+ \alpha' \partial_\tau X^- -g_{\sigma\sigma} \sum_{i=1}^8 \left((\partial_\tau X^i)^2 + \frac{(\alpha' p^+)^2}{\epsilon^2}\Omega(\alpha' p^+ \tau/\epsilon) (X^i)^2\right)\\
&-2 g_{\tau\sigma}\left(\alpha' p^+ \partial_\sigma X^- - \partial_\tau X^i \partial_\sigma X^i\right)+g_{\sigma\sigma}^{-1}(1-g_{\tau\sigma}^2)\sum_{i=1}^8 (\partial_\sigma X^i)^2-\frac{1}{2}\alpha' R^{(2)}\Phi\Biggr).\nonumber
\end{align}
We rescale $\epsilon=\epsilon' \alpha' p^+$,
\be
\frac{(\alpha' p^+)^2}{\epsilon^2}\Omega(X^+/\epsilon)=\frac{1}{\epsilon'^2}\Omega(\tau/\epsilon'),
\ee
and from here on, we will denote worldsheet time $\tau=t$ and write $\epsilon$ instead of $\epsilon'$. The $\sigma$-dependent part of the oscillator $X^-$ is non-dynamical and enforces $g_{\tau\sigma}=0$. The $\sigma$-independent part of the oscillator $X^-$ can be eliminated as a constraint ($g_{\sigma\sigma}=1$), the (dynamically non-trivial) coupling to the dilaton disappears (see, e.g., \cite{PRT}), and we can write the following worldsheet Hamiltonian
\be
H=\frac{1}{4\pi\alpha'} \int \mathrm{d}\sigma  \sum_{i=1}^d \left(\pi^2  (P_i)^2 + \frac{\lambda}{\epsilon^2}\Omega(\tau/\epsilon) (X^i)^2 + \left(\partial_\sigma X^i\right)^2\right),\label{lcHamiltonian}
\ee
where $P_i$ are momenta conjugate to $X^i$. We will now choose units in which $\alpha'=1$. If we Fourier transform the $\sigma$-coordinate, 
\be
X^i(t,\sigma)=X^i_0(t)+ \sqrt{2}\sum_{n>0} \left(\mathrm{cos}\left(n\sigma\right) X^i_{n}(t)+\mathrm{sin}\left(n\sigma\right) \tilde{X}^i_{n}(t)\right),
\ee
we obtain a set of time-dependent harmonic oscillator Hamiltonians
\begin{align}
H&= \sum_{n=0}^\infty \sum_{i=1}^d H_{ni},\label{setham}\\
H_{0i}&=\frac{(P_{0i})^2}{2}+ \frac{\lambda}{\epsilon^2}\Omega(t/\epsilon) \frac{(X_{0}^i)^2}{2},\\
H_{ni}&=\frac{(P_{ni})^2+(\tilde{P}_{ni})^2}{2}+ \left(n^2 + \frac{\lambda}{\epsilon^2}\Omega(t/\epsilon)\right) \frac{(X_{n}^i)^2+  (\tilde{X}_{n}^i)^2}{2}.\label{HOham}
\end{align}

\subsection{WKB solution for time-dependent harmonic oscillator}

The Hamiltonian (\ref{setham}) is quadratic and the solution to the corresponding Schr\"odinger equation,
\be
i\frac{\partial}{\partial t} \phi(t;X^i_n)=\left(\sum_{n} \sum_{i=1}^d H_{ni}\right) \phi(t;X^i_n),\label{schr_eq}\\
\ee
can be found using WKB techniques, which are exact for quadratic Hamiltonians. From (\ref{schr_eq}) it follows that
\be
i\frac{\partial}{\partial t} \phi_n^i(t;X^i_n)=-\frac{1}{2}\frac{\partial^2}{(\partial X_{n}^i)^2} \phi_n^i(t;X^i_n)+ \frac{1}{2} \left(n^2 + \frac{\lambda}{\epsilon^2}\Omega(t/\epsilon)\right)\left(X_{n}^i\right)^2 \phi_n^i(t;X^i_n),
\ee
if we separate variables as
\be
\phi(t;\textbf{X})=\prod_n\prod_{i=1}^8 \phi_n^i(t;X^i_n).
\ee
We then take the WKB ansatz
\be
\phi_n^i(t;X)=\mathcal{A}_n(t_1,t)\,\mathrm{exp}\left(i S_{cl;n}[X_{1,n}^i,t_1|X_n^i,t]\right),
\label{WKB}
\ee
where $S_{cl;n}[X_{1,n}^i,t_1|X_n^i,t]$ is the ``classical action'' evaluated for the path going from $X_{1,n}^i$ at the time $t_1$ to $X_n^i$ at the time $t$,
\be
S_{cl}[X_{1,n}^i,t_1|X_n^i,t]=\int_{t_1}^{t} \mathrm{d}t' \left( \frac{(\dot{X_n^i})^2}{2}-\left(n^2+\frac{\lambda}{\epsilon^2}\Omega\left(\frac{t'}{\epsilon}\right)\right)\frac{(X_n^i)^2}{2}\right).
\ee
If $\mathcal{A}_n(t_1,t)$ satisfies
\be
-2\frac{\partial}{\partial t}\mathcal{A}_n(t_1,t)=\mathcal{A}_n(t_1,t)\frac{\partial^2}{\partial (X_n^i)^2} S_{cl}[X_{1,n}^i,t_1|X_n^i,t],
\ee
then (\ref{WKB}) satisfies the original Schr\"odinger equation exactly.

Up to normalization, a basis of solutions, labelled by the initial condition $X_n^i(t_1)=X_{1,n}^i$, is given by \cite{EvninNguyen},
\be
\phi(t;X_n^i) \sim \prod_{ni}\frac{1}{\sqrt{\mathcal{C}(t_1,t)}}\mathrm{exp}\left(-\frac{i}{2 \mathcal{C}}\sum_{i=1}^d \left[(X_{1,n}^i)^2\partial_{t_1}\mathcal{C}-(X_n^i)^2\partial_{t_2}\mathcal{C}+2 X_{1,n}^i X_n^i\right]\right),\label{basisofsols}
\ee
where $\mathcal{C}(t_1,t_2)$ (suppressing the index $n$) is a solution to the ``classical equation of motion'' for the time-dependent harmonic oscillator Hamiltonian (\ref{HOham}):
\be
\partial^2_{t_2}{\mathcal{C}}(t_1,t_2)+\left(n^2+\frac{\lambda}{\epsilon^2}\,\Omega(t_2/\epsilon)\right)\mathcal{C}(t_1,t_2)=0,\label{ceomC}
\ee
with initial conditions specified as
\be
\mathcal{C}(t_1,t_2)|_{t_1=t_2}=0,\hspace{5mm}\partial_{t_2} \mathcal{C}(t_1,t_2)|_{t_1=t_2}=1.\label{initC}
\ee
We shall refer to $\mathcal{C}(t_1,t_2)$ as ``compression factor'', since it describes
convergence of solutions to the corresponding harmonic oscillator equation starting
at the same point at the moment $t_1$. (If $\mathcal{C}(t_1,t_2)$ vanishes, then
$t_2$ is a focal point, as the difference between any two solutions with the same initial position $X(t_1)$ is proportional to $\mathcal{C}(t_1,t_2)$.) A useful representation of $\mathcal{C}(t_1,t_2)$ is given by
\be
\mathcal{C}(t_1,t_2)=\frac{f(t_1) h(t_2)-f(t_2)h(t_1)}{W[f,h]},\label{repCfh}
\ee
where $f(t)$ and $h(t)$ are two independent solutions to the differential equation under consideration, and the Wronskian $W$ is given by
\be
W[f,h]=f\dot{h}-h\dot{f}.\label{wronskian}
\ee
To derive the singular limit of the wavefunction (\ref{basisofsols}) it is sufficient to study the singular limit of (\ref{ceomC}-\ref{initC}).

%%%%%%%%%%%%%%%%%%%%%%%%%%%%%%%%%%%%%%%%%%%%%%%%

\section{The singular limit for the center-of-mass motion}
\label{0mode}

For the $n=0$ mode, we obtain as the ``classical equation of motion''
\be
\ddot{X}+\frac{\lambda}{\epsilon^2}\Omega(t/\epsilon)X=0.
\ee
We need to study the $\epsilon\rightarrow 0$ limit of the solution that obeys the initial conditions
\be
X(t_1)=0,\hspace{10mm}\dot{X}(t_1)=1,\hspace{10mm}t_1<0. 
\label{initcondX}
\ee

The singular limit of solutions to this equation has been analyzed in \cite{EvninNguyen}. Performing a scale transformation $Y(\eta)=X(\eta\epsilon)$, with $\eta=t/\epsilon$, removes the $\epsilon$-dependence from the equation, leaving
\be
\frac{\partial^2}{\partial \eta^2}{Y}+\lambda\Omega(\eta)Y=0.
\label{etaEq}
\ee
This scale transformation is possible because our initial singular metric was scale-invariant and we have resolved it as in (\ref{scaleinvres}) without
introducing any dimensionful parameters besides $\epsilon$. The existence of a singular limit is then translated \cite{EvninNguyen} into constraints on on the asymptotic behavior of solutions to (\ref{etaEq}). These ``boundary conditions at infinity'' are strongly reminiscent of a Sturm-Liouville problem, and it is natural 
that a discrete spectrum of $\lambda$ is singled out by imposing the existence
of a singular limit.

For the specific asymptotics of our resolved profile (\ref{asrefprof}), it can be shown \cite{EvninNguyen}
that, in the infinite past and infinite future, the solutions approach a linear combination of two powers (denoted below $a$ and $1-a$, with $a$ being a function
of $k\lambda$, cf. (\ref{scaleinvres}-\ref{asrefprof})). This power law behavior simply corresponds to the regime when
the second term on the right hand side of (\ref{asrefprof}) can be neglected compared to the first. It is then convenient to form two bases of solutions, one asymptotically approaching the two powers (dominant and subdominant) at $\eta\rightarrow -\infty$,
\be
Y_{1-}(\eta)=|\eta|^{a_-} + o(|\eta|^{a_-}),\hspace{10mm}Y_{2-}(\eta)=|\eta|^{1-a_-} + o(|\eta|^{1-a_-}),
\label{Y-}
\ee
and another behaving similarly at $\eta\rightarrow +\infty$
\be
Y_{1+}(\eta)=|\eta|^{a_+} + o(|\eta|^{a_+}),\hspace{10mm}Y_{2+}(\eta)=|\eta|^{1-a_+} + o(|\eta|^{1-a_+}),
\label{Y+}
\ee
where $a_\pm$ is given by
\be
a_\pm=\frac12+\sqrt{\frac14-\lambda k_\pm}.
\ee
(We are temporarily assuming that $k$ can take two different values $k_\pm$ for
the positive and negative time asymptotics, a possibility that will be discarded shortly.) The two bases need, of course, to be related by a linear transformation:
\be
\begin{bmatrix}Y_{1-}(\eta)\\Y_{2-}(\eta)\end{bmatrix}=Q(\lambda)\begin{bmatrix}Y_{1+}(\eta)\\Y_{2+}(\eta)\end{bmatrix},
\label{Q}
\ee
where $Q(\lambda)$ is a $2\times 2$ matrix whose determinant is constrained by Wronskian conservation as
\be
W[Y_{1-},Y_{2-}]=W[Y_{1+},Y_{2+}]\det Q .
\ee

The singular limit has been rigorously considered in \cite{EvninNguyen},
but the results can be understood heuristically from the following argument.
Imagine one is trying to construct a solution $\tilde Y$ to (\ref{etaEq}) satisfying
some ($\epsilon$-independent) initial conditions at $\eta_1=t_1/\epsilon<0$. This solution can be
expressed in terms of $Y_{1-}$ and $Y_{2-}$ (a complete basis) as
\be
\tilde Y=C_1 Y_{1-}+C_2 Y_{2-}.
\ee
Since the initial conditions are specified at $\eta_1=t_1/\epsilon$, the asymptotic
expansions (\ref{Y-}) are valid. There needs to be a non-trivial contribution
from both $Y_{1-}$ and $Y_{2-}$ in the above formula in order to satisfy general
initial conditions. Hence, the two terms on the right hand side should be of order 1.
Therefore, we should have
\be
C_1=O(\epsilon^{a_-}),\qquad C_2=O(\epsilon^{1-a_-}).
\ee
If we now apply (\ref{Q}) and (\ref{Y+}) to evaluate $\tilde Y$ at a large positive $\eta=t_2/\epsilon$, the powers of $\epsilon$ in $C_1$ and $C_2$ will combine
with the powers of $\epsilon$ originating from $Y_{1+}$ and $Y_{2+}$ and yield
\be
\begin{array}{l}
\dsty\tilde Y(t_2/\epsilon)=Q_{11}(\lambda)t_2^{a_+}O(\epsilon^{a_--a_+})+Q_{12}(\lambda)t_2^{1-a_+}O(\epsilon^{a_-+a_+-1})\vspace{2mm}\\
\dsty\hspace{3cm}+Q_{21}(\lambda)t_2^{a_+}O(\epsilon^{1-a_--a_+})+Q_{22}(\lambda)t_2^{1-a_+}O(\epsilon^{a_+-a_-}).
\end{array}
\ee
Since $a_+$ and $a_-$ are greater than $1/2$, this expression can only have an $\epsilon\to 0$ limit if $a_+=a_-$ (i.e., $k_+=k_-$ and we can set both
equal to 1 by redefining $\lambda$) and $Q_{21}(\lambda)=0$.
The latter condition implies that the absolute normalization $\lambda$ of the plane wave profile $\Omega(\eta)$ will generically lie in a discrete spectrum, dependent on the specific way the singularity is resolved, i.e., the shape of $\Omega(\eta)$. A particular exactly solvable example for this discrete spectrum (there called ``light-like reflector plane'') has been given in \cite{CDE}. With $Q_{21}(\lambda)=0$ and $\det Q=-1$, the matrix $Q$ can be written as
\be
Q=\begin{bmatrix}q&\tilde q\\0&-1/q\end{bmatrix},
\label{asbas}
\ee
with $q$ being a real nonzero number ($\tilde q$ does not affect the singular limit). For flat space-time we have $q=1$ and for the ``light-like reflector plane'' of \cite{CDE} we have $q=-1$. In the singular limit, a basis of solutions is given by
\begin{align}
&Y_{1}(t)=(-t)^a,\hspace{5mm}Y_{2}(t)=(-t)^{1-a},\hspace{5mm}t<0,\nonumber\\
&Y_{1}(t)=q\,t^a,\hspace{5mm}Y_{2}(t)=-\frac{1}{q}t^{1-a},\hspace{5mm}t>0.\label{basissolzeromode}
\end{align}
%%%%%%%%%%%%%%%%%%%%%%%%%%%%%%%%%%%%%%%%

\section{The singular limit for excited string modes}

Following our general discussion of free strings in plane wave backgrounds,
the evolution of excited string modes is described by time-dependent harmonic oscillator equations
\be
\frac{\partial^2}{\partial t^2} X(t) + \left(n^2+\frac{\lambda}{\epsilon^2}\Omega(t/\epsilon)\right) X(t)=0.\label{de}
\ee
Solutions for the wavefunctions of the excited string modes can be expressed in
terms of a particular solution to this equation $\mathcal{C}(t_1,t_2)$ defined by (\ref{ceomC}-\ref{initC}). Hence, to analyze the singular ($\epsilon\to 0$) limit of the excited modes dynamics, it should suffice to analyze the singular limit of $\mathcal{C}(t_1,t_2)$. Because $n^2$ is finite, it is natural to expect that it does not affect the existence of the singular limit (governed by the singularity emerging from $\Omega(t/\epsilon)$). We shall prove that it is indeed the case for positive $\lambda$ (for negative $\lambda$ unstable motion of the inverted harmonic oscillator
leads to divergences\footnote{More specifically, the divergences arise from sub-leading infinities in the position of the inverted harmonic oscillator, while the leading infinities cancel. (Such sub-leading infinities are absent for the center-of-mass motion analyzed in section \ref{0mode}, hence no analogous divergences in that case.) Further details are given in section 4.4 and appendix A.}).

To derive $\mathcal{C}(t_1,t_2)$ for equation (\ref{de}) we use the following strategy: the differential equation (\ref{de}) is linear and any solution $X(t_2)$ at $t=t_2$ can be written in terms of a ``transfer matrix'' $T$ that only depends on the initial and final times,
\be
\begin{bmatrix} X(t_2)\\\dot X(t_2)\end{bmatrix}=T(t_1,t_2)\begin{bmatrix} X(t_1)\\\dot X(t_1)\end{bmatrix}.
\ee
The transfer matrix can be expressed as
\be
T(t_1,t_2)=\begin{bmatrix}-\partial_{t_i}\mathcal{C}(t_1,t_2)&\mathcal{C}(t_1,t_2)\\- \partial_{t_i}\partial_{t_f}\mathcal{C}(t_1,t_2)&\partial_{t_f} \mathcal{C}(t_1,t_2)\end{bmatrix},\label{TifoC}
\ee
where $\partial_{t_i}$ and $\partial_{t_f}$ indicate differentiation with respect to the first and second argument respectively. The transfer matrix is completely determined once $\mathcal{C}(t_1,t_2)$ has been determined, and vice versa. We will now use the fact that transfer matrices of subintervals are combined by ordinary
matrix multiplication. Dividing the solution region into three sub-intervals, we shall calculate the transfer matrices $T_k$ for each sub-interval $k$ and apply multiplication to construct the total transfer matrix. The sub-intervals shall be chosen as indicated in the following figure:

\setlength{\unitlength}{1mm}
\noindent\begin{picture}(160,20)
\put(0,10){\line(1,0){160}}
\put(40,15){\text{I}}
\put(79,15){\text{II}}
\put(120,15){\text{III}}
\put(15,9){\text{$|$}}
\put(65,9){\text{$|$}}
\put(95,9){\text{$|$}}
\put(145,9){\text{$|$}}
\put(15,4){\text{$t_1$}}
\put(63,4){\text{$-t_\epsilon$}}
\put(95,4){\text{$t_\epsilon$}}
\put(145,4){\text{$t_2$}}
\put(80,9){\text{$|$}}
\put(80,4){\text{$0$}}
\end{picture}

\noindent We use $t_\epsilon$ to indicate a time that will approach zero in the singular limit as 
\be
t_\epsilon=\epsilon^{1-c}\tilde t^c,
\label{teps}
\ee
with $\tilde t$ staying finite in relation to the ``moments of observation'' $t_1$ and $t_2$. The number $c$ (between 0 and 1) will be chosen later as needed for our proof. On each interval, we can write the transfer matrix $T_k$ in terms of the ``compression factor'' $\mathcal{C}_k$. Using matrix multiplication to construct the full transfer matrix $T$, we can now deduce an expression for the ``compression factor'' of the complete interval
\begin{align}
\mathcal{C}(t_1,t_2)=&\mathcal{C}_{I}(t_1,-t_\epsilon)\partial_{t_i}\mathcal{C}_{II}(-t_\epsilon,t_\epsilon)\partial_{t_i}\mathcal{C}_{III}(t_\epsilon,t_2)-\partial_{t_f}\mathcal{C}_{I}(t_1,-t_\epsilon)\mathcal{C}_{II}(-t_\epsilon,t_\epsilon)\partial_{t_i}\mathcal{C}_{III}(t_\epsilon,t_2)\nonumber\\&-\mathcal{C}_{I}(t_1,-t_\epsilon)\partial_{t_i}\partial_{t_f}\mathcal{C}_{II}(-t_\epsilon,t_\epsilon)\mathcal{C}_{III}(t_\epsilon,t_2)+\partial_{t_f}\mathcal{C}_{I}(t_1,-t_\epsilon)\partial_{t_f}\mathcal{C}_{II}(-t_\epsilon,t_\epsilon)\mathcal{C}_{III}(t_\epsilon,t_2),
\label{Cexpr0}
\end{align}
in terms of the ``compression factors'' of the three sub-intervals. Once again, $\partial_{t_i}$ and $\partial_{t_f}$ differentiate $\mathcal{C}$ with respect to its first and second argument (initial and final time).

To study the existence of the singular limit of $\mathcal{C}(t_1,t_2)$, we will use the following strategy: for two linear differential equations related by a small perturbation we will establish a bound on the difference between perturbed and unperturbed solutions with the same initial conditions. This bound will, of course,
apply to $\mathcal{C}_k$. For each of the three sub-intervals introduced above, we will consider a simplified differential equation that is a good approximation to equation (\ref{de}) on the corresponding interval:
\begin{itemize}
\item Region I and III: $\ddot{X}(t) + \left(n^2+ \lambda/t^2\right) X(t)=0$ (related to Bessel's equation);
\item Region II: $\ddot{X}(t) + \lambda/\epsilon^2 \Omega(t/\epsilon) X(t)=0$ (equation of motion for the zero mode).
\end{itemize}
Then, on each sub-interval, $\mathcal{C}_k$ can be written as the sum of a simplified ``compression factor'' $\bar{\mathcal{C}}_k$ satisfying the simplified differential equation on this sub-interval, plus a small perturbation $\delta \mathcal{C}_k$. We will prove that, in the singular limit, the $\delta \mathcal{C}_k$ will drop out of the expression for the total ``compression factor'' $\mathcal{C}(t_1,t_2)$.

Most of this section is dedicated to implementing the proof we have just outlined. The reader primarily interested in the discussion of the singular limit and content with
the general sketch given above can skip to section \ref{eff_mat_cond}.

\subsection{Bounds on solutions to perturbed differential equations}

In view of the subsequent application to the singular limit analysis, we would like to bound the difference $\delta X$ between the solution $X(t)$ of a perturbed differential equation,
\be
\frac{\partial^2}{\partial t^2} X + \left(\Upsilon+\delta\Upsilon\right) X=0,\label{pert_de}
\ee
and the solution $\bar{X}(t)$ of an unperturbed differential equation,
\be
\frac{\partial^2}{\partial t^2} \bar{X} + \Upsilon \bar{X}=0, \label{orig_de}
\ee
where we take 
\be
X=\bar{X}+\delta X\label{partX}
\ee
and demand that the initial conditions remain unchanged:
\be
X(t_0)=\bar{X}(t_0),\hspace{10mm} \partial_t X(t_0)=\partial_t\bar{X}(t_0).\label{IC}
\ee
If we substitute (\ref{partX}) and (\ref{orig_de}) into (\ref{pert_de}) we obtain a differential equation for the perturbation on the solution.
\begin{equation}
\frac{\partial^2}{\partial t^2} \delta X +\Upsilon(t) \delta X  = - \delta\Upsilon(t) \left(\bar{X}+\delta X\right).\label{rest_de} 
\end{equation}
A formal solution to (\ref{rest_de}) is given by
\be
\delta X (t)  = - \int_{-\infty}^\infty  G_r(t,t') \delta\Upsilon(t') \left(\bar{X}(t')+\delta X(t')\right) dt', \label{formalSol}
\ee
with the Green function $G_r(t,t')$ satisfying
\be
\left(\frac{\partial^2}{\partial t^2} + \Upsilon(t) \right) G_r(t,t')=\delta(t-t')
\ee
and initial conditions
\be
G_r(t,t')|_{t=t_0}=0,\hspace{10mm}\partial_t G_r(t,t')|_{t=t_0}=0.
\ee
Therefore, we can write the Green function in terms of the ``compression factor'' $\mathcal{\bar{C}}$ of the unperturbed equation (\ref{orig_de}), where $\mathcal{\bar{C}}$ obeys the same initial conditions as in (\ref{initC}):
\be
G_r(t,t')=\begin{cases} \mathcal{\bar{C}}(t',t) & t_0<t'<t,\\ 0&\mathrm{otherwise}.  \end{cases}
\label{GrC}
\ee
To obtain a bound on $\delta X$ we will invoke the so-called Gronwall inequality \cite{gronwall}.\\

\subsubsection{The Gronwall inequality}
\emph{Let $I=[A,B]$. Assume $\beta$ and $\alpha$ real valued and continuous on $I$ and $\beta\geq0$. If $u$ is continuous, real valued on I and satisfies the integral inequality
\be
u(t)<\alpha(t)+\int_A^t \beta(s) u(s) ds,\hspace{10mm}t\in I,\label{assump}
\ee
then
\be
u(t)<\alpha(t)+\int_A^t  \beta(s) \alpha(s) \; \mathrm{exp}\left(\int_s^t \beta(r)dr\right) ds, \hspace{10mm}t\in I.
\ee}
\textbf{Proof}: First we define
\be
z(t)=\int_A^t \beta(s)u(s)\mathrm{d}s,\hspace{10mm}t\in I.
\ee
Then, after differentiation and using the initial assumption (\ref{assump}), we obtain
\be
z'(t)=\beta(t)u(t)\leq \beta(t) \alpha(t)+ \beta(t) z(t).
\ee
Using the line above we write
\begin{align}
\Bigl[\mathrm{exp}\left(-\int_A^s \beta(u)\mathrm{d}u\right)z(s)\Bigr]'&=\mathrm{exp}\left(-\int_A^s \beta(r)\mathrm{d}r\right)\left(z'(s)-\beta(s)z(s)\right)\\
&\leq \beta(s)\alpha(s) \mathrm{exp}\left(-\int_A^s \beta(u)\mathrm{d}u\right)\hspace{10mm}s\in I.
\end{align}
We integrate from $a$ to $t$ and obtain,
\be
\mathrm{exp}\left(-\int_A^t \beta(s)\mathrm{d}s\right)z(t)\leq \int_A^t \beta(s)\alpha(s) \mathrm{exp}\left(-\int_A^s \beta(u)\mathrm{d}u\right) \mathrm{d}s\hspace{10mm}t\in I.\label{tussen}
\ee
From assumption (\ref{assump}) and (\ref{tussen}) we now derive the desired inequality,
\begin{align}
u(t)\leq \alpha(t) + z(t) &\leq \alpha(t) + \mathrm{exp} \left( \int_A^t \beta(r) \mathrm{d}r \right) \int_A^t \beta(s)\alpha(s) \mathrm{exp} \left( -\int_A^s \beta(u)\mathrm{d}u \right) \mathrm{d}s \\
&=\alpha(t) + \int_A^t \beta(s)\alpha(s) \mathrm{exp}\left(\int_s^t \beta(u)\mathrm{d}u\right) \mathrm{d}s,\hspace{10mm}t\in I.
\end{align}

\subsubsection{Bounds on the perturbations $\delta X$}

From (\ref{formalSol}) we derive the following bound on the formal solution $\delta X$ 
\be
|\delta X (t)|
<\int_{-\infty}^\infty  |G_r(t,t') \delta\Upsilon(t') \bar{X}(t')| dt' + \int_{-\infty}^\infty |G_r(t,t') \delta\Upsilon(t') \delta X(t')| dt'. 
\ee
We will now use the fact that, by virtue of (\ref{GrC}), where nonzero, $G_r(t,t')=C(t',t)$. Hence (cf. (\ref{repCfh})), inside the integral,
\be
|G(t,t')|<\frac{1}{|W|}\left(|f|_M|h(t')|+|f(t')||h|_M\right)\equiv g(t'),
\label{greenbound}
\ee
with $|f|_M$ and $|h|_M$ being the absolute value maxima of these functions on the integration domain. The integration regions are in fact finite, since (\ref{GrC}) vanishes unless $t_0<t'<t$:
\be
|\delta X (t)| <\int_{t_0}^t  |g(t') \delta\Upsilon(t') \bar{X}(t')| dt' + \int_{t_0}^t |g(t') \delta\Upsilon(t') \delta X(t')| dt'.
\ee
Since $g(t')$ is independent of $t$ we can now apply Gronwall's inequality to obtain
\begin{multline}
|\delta X (t)| <\int_{t_0}^t  |g(t') \delta\Upsilon(t') \bar{X}(t')| dt' \\+ \int_{t_0}^t \left(\int_{t_0}^{t'}  |g(t'') \delta\Upsilon(t'') \bar{X}(t'')| dt'' \right)| g(t') \delta\Upsilon(t')| \mathrm{exp}\left(\int_{t'}^t | g(t'') \delta\Upsilon(t'')|  dt''\right) dt'.
\end{multline}
On the interval $(t_0,t)$ we assume the existence of a maximum of $|\bar X|$ and of $|\delta\Upsilon|$ and we call these $|\bar{X}|_M$ and $|\delta\Upsilon|_M$ respectively. We also assume the integral $\int_{t_0}^t  |g(t')| dt'$ can be bounded by a number $M$. If
\be \int_{t_0}^t  |g(t')| dt'<M,
\label{Mbound}
\ee
then it follows that also
\be
\int_{t'}^t  |g(t'')| dt''<M.
\ee
We thus find
\be
|\delta X (t)| < |\bar{X}|_M \left(M |\delta\Upsilon|_M + M^2 |\delta\Upsilon|_M^2 \mathrm{exp}\left(M |\delta\Upsilon|_M\right)\right).
\label{grnwl}
\ee
The second term on the right-hand side is negligible compared to the first one
for sufficiently small $|\delta\Upsilon|$.

\subsection{Solutions away from the singularity\label{regIandIII}}

In regions I and III we will take
\be
\Upsilon=n^2+k/t^2,\hspace{10mm}\delta \Upsilon=\frac1{\epsilon^2}\,O\left(\frac{\epsilon^b}{t^b}\right),
\ee
with $b$ defined in (\ref{asrefprof}).
The solutions to the unperturbed differential equation (\ref{orig_de}) are given by 
\be
\sqrt{|t|}J_\alpha(|nt|),\hspace{10mm}\sqrt{|t|}J_{-\alpha}(|nt|),\hspace{10mm}\alpha=a-\frac{1}{2},\label{solsregI}
\ee
where the Bessel functions, $J_{\alpha}(x)$ and $J_{-\alpha}(x)$, satisfy the differential equation
\be
x^2\frac{\partial^2}{\partial x^2}J_\alpha(x)+x\frac{\partial}{\partial x}J_\alpha + \left(x^2-\alpha^2\right)J_\alpha(x)=0.
\ee
(This Bessel-negative-order-Bessel basis is more convenient for our purposes
than the often-used Bessel-Neumann basis, as it approaches $|t|^a$ and $|t|^{1-a}$ for small values of $t$ without mixing the two powers.)

The unperturbed ``compression factor'' in region I is then 
\be
\bar{\mathcal{C}}_{I}(t_1,t)=\sqrt{|t_1|}\sqrt{|t|}\frac{J_\alpha(-nt_1)J_{-\alpha}(-nt)-J_\alpha(-nt)J_{-\alpha}(-nt_1)}{W[\sqrt{|t|} J_\alpha(-nt),\sqrt{|t|} J_ {-\alpha}(-nt)]}.
\label{CregI}
\ee
Using the series expansion of the Bessel function for small arguments (they will be evaluated at $t=-t_\epsilon$),
\be
J_\alpha(x)\sim \left(\frac{x}{2}\right)^\alpha\frac{1}{\Gamma(\alpha+1)},\hspace{10mm}\alpha\neq-1,-2,-3,\ldots.\label{seriesExpBessel}
\ee
we can estimate the various contributions to (\ref{grnwl}), thereby constraining the correction to the unperturbed ``compression factor''. One can distinguish three cases:\vspace{2mm}

\noindent 1) $a>1, J_\alpha(-nt_1)\ne 0$, which yields
\be
|\bar{\mathcal{C}}(t_1,t_\epsilon)|\propto\epsilon^{(1-c)(1-a)},\hspace{5mm}|\bar{\mathcal{C}}|_M\propto\epsilon^{(1-c)(1-a)},\hspace{5mm}|\delta\Upsilon|_M\propto\epsilon^{bc-2},\hspace{5mm}M\propto\epsilon^{(1-c)(1-a)}.
\ee
From (\ref{grnwl}), $\delta\mathcal{C}(t_1,t_\epsilon)$ is negligible compared to $\mathcal{C}(t_1,t_\epsilon)$ if
\be
c>\frac{a+1}{a+b-1}.
\ee

\noindent 2) $a<1, J_\alpha(-nt_1)\ne 0$, which yields
\be
|\bar{\mathcal{C}}(t_1,t_\epsilon)|\propto\epsilon^{(1-c)(1-a)},\hspace{5mm}|\bar{\mathcal{C}}|_M\propto\epsilon^{0},\hspace{5mm}|\delta\Upsilon|_M\propto\epsilon^{bc-2},\hspace{5mm}M\propto\epsilon^{0}.
\ee
From (\ref{grnwl}), $\delta\mathcal{C}(t_1,t_\epsilon)$ is negligible compared to $\mathcal{C}(t_1,t_\epsilon)$ if
\be
c>\frac{3-a}{b+1-a}.
\ee

\noindent 3) $J_\alpha(-nt_1)=0$, which yields
\be
|\bar{\mathcal{C}}(t_1,t_\epsilon)|\propto\epsilon^{(1-c)a},\hspace{5mm}|\bar{\mathcal{C}}|_M\propto\epsilon^{0},\hspace{5mm}|\delta\Upsilon|_M\propto\epsilon^{bc-2},\hspace{5mm}M\propto\epsilon^{0}.
\ee
From (\ref{grnwl}), $\delta\mathcal{C}(t_1,t_\epsilon)$ is negligible compared to $\mathcal{C}(t_1,t_\epsilon)$ if
\be
c>\frac{2+a}{b+a}.
\ee

Whichever of the three cases is realized, it suffices for $c$ to be greater than
a number less than 1, in order for the corrections to the unperturbed ``compression factor'' to be negligible for small values of $\epsilon$. The discussion of interval III is completely parallel to what we have just presented.

\subsection{Solutions in the near-singular region\label{regII}}

The ``unperturbed'' equation in region II,
\be
\frac{\partial^2}{\partial t^2}\bar{X}(t)+\frac{\lambda}{\epsilon^2}\Omega(t/\epsilon)\bar{X}(t)=0,
\ee
is precisely that of the string center-of-mass motion. In order to simplify derivations, we shall assume $a_+=a_-$,
as required for well-defined zero-mode propagation (see section \ref{0mode}). The unperturbed ``compression factor'' in region II takes the form\footnote{This formula follows from (\ref{Y-}), (\ref{Y+}) and (\ref{Q}) via (\ref{repCfh}). It is also the same as (44) of \cite{EvninNguyen}. The expression given corresponds to small values of $\epsilon$. The corrections are suppressed by powers of $\epsilon$ and do not contribute to the singular limit.}
\be
\bar{\mathcal{C}}_{II}(t_i,t_f)=\frac{Q_{22}(\lambda) |t_i|^a t_f^{1-a} -Q_{11}(\lambda)|t_i|^{1-a} t_f^a - Q_{12}(\lambda)|t_i|^{1-a} t_f^{1-a}\epsilon^{2a-1} +Q_{21}(\lambda) |t_i|^a t_f^a\epsilon^{1-2a}}{2a-1},
\label{CII}
\ee
where the $2\times 2$ matrix $Q$ is defined by (\ref{asbas}), and we have used $W[|t|^a,|t|^{1-a}]=2a-1$. To study the perturbation we will first perform the scaling transformation $\eta=t/\epsilon$, $Y(\eta)=X(\eta\epsilon)$, which yields
\be
\frac{\partial^2}{\partial \eta^2} Y(\eta) + \left(\epsilon^2 n^2+\lambda \Omega(\eta)\right) Y(\eta)=0.\label{scaleinv_de}
\ee
We now take
\be
\Upsilon=\lambda\Omega(\eta),\hspace{10mm}\delta \Upsilon=\epsilon^2 n^2.
\label{n2}
\ee
If we now choose $f\sim\eta^a$, $g\sim\eta^{1-a}$ in (\ref{greenbound}), $M$ of (\ref{Mbound}) for the region $(-t_\epsilon/\epsilon,t_\epsilon/\epsilon)$ (whose size, in $\eta$,
is proportional to $\epsilon^{-c}$) becomes (with the three factors coming from $f$, $h$ and the size of the integration region):
\be
M\propto\epsilon^{-ac}\epsilon^{-(1-a)c}\epsilon^{-c}=\epsilon^{-2c}.
\ee
Because there are only power laws involved in (\ref{CII}), the maximal value
$\bar{\mathcal{C}}_M$ is of the same order as $|\bar{\mathcal{C}}(-t_\epsilon,t_\epsilon)|$. Furthermore, $|\delta\Upsilon|_M\propto\epsilon^{2}$ by construction. It then follows from (\ref{grnwl}) that
\be
|\delta \mathcal{C}_{II}| < \left( O\left(\epsilon^{2-2c}\right) +  O\left(\epsilon^{4-4c} \mathrm{exp}\left(\epsilon^{2-2c}\right)\right)\right)|\bar{\mathcal{C}}_{II}|.
\ee
The correction is negligible for any $c<1$.

A subtlety in our above derivation deserves a comment (we would like to thank the JHEP referee for raising this point): one might have thought
that the factor of $n^2$ in $\delta\Upsilon$ of (\ref{n2}) competes with
the smallness of $\epsilon$ and undermines the validity of our considerations
(for sufficiently large mode numbers). It is indeed true that, for each value
of $\epsilon$ (each fixed resolved space), our analysis is only valid 
for modes with sufficiently small mode numbers (though this range of validity
increases infinitely as $\epsilon$ is taken to 0). However, since the modes
are completely independent, the limit for the motion of the entire string
(if it exists) is exactly the same as if it were computed mode-by-mode.
For that reason, $n$ can be kept fixed in the derivations of this section,
and the problem of $n^2$ competing with the smallness of $\epsilon$ does not
arise. (This attitude guarantees reproducing the $\epsilon\to 0$ limit
correctly for the entire set of modes, though it does not allow to draw
conclusions on the uniformity of this limit with respect to $n$.)

\subsection{Effective matching conditions \label{eff_mat_cond}}

Having analyzed the ``compression factors'' on subintervals I, II and III, we can combine them into the total ``compression factor'' by applying (\ref{Cexpr0}).
As has been shown above, there exist a number $c$ in (\ref{Cexpr0}) between
0 and 1, such that the ``compression factors'' on subintervals I, II and III
can be well approximated by the simplified expressions (\ref{CregI}) and (\ref{CII}),
with corrections suppressed by positive powers of $\epsilon$. One can then
substitute (\ref{CregI}) and (\ref{CII}) into the right-hand-side of (\ref{Cexpr0}).

For $a>1$ ($\lambda k<0$), the Bessel functions featured in (\ref{CregI}) blow up
near the origin (the inverted harmonic oscillator is
propelled off to infinity). This threatens the existence of an $\eps\to 0$ limit.
In appendix A, we display the divergences arising for $a>3/2$. (For $1<a<3/2$, the limit may exist for individual string modes, but a consideration along the lines of section \ref{total} would still indicate no consistent propagation for the entire string.) 
In any case, we shall not explore this case further since, as will be explained in section \ref{discussion}, free strings are not a good approximation to motion in such plane waves.

For $a<1$ ($\lambda k>0$), substituting (\ref{CregI}) and (\ref{CII}) in (\ref{Cexpr0}) yields
\be
\begin{array}{l}
\dsty\bar{\mathcal{C}}(t_1,t_2)=\frac{\sqrt{-\pi t_1 t_2}}{2\sin\alpha\pi}\Bigl(Q_{22}(\lambda) J_{a-1/2}(-n t_1)J_{1/2-a}(n t_2)-Q_{11}(\lambda) J_{1/2-a}(-n t_1)J_{a-1/2}(n t_2) \vspace{2mm}\\ 
\hspace{3cm}+Q_{21}(\lambda) \epsilon^{1-2a} J_{a-1/2}(-n t_1)J_{a-1/2}(n t_2) \gamma_n\vspace{2mm}\\
\hspace{3cm}- Q_{12}(\lambda) \epsilon^{2a-1}J_{1/2-a}(-n t_1)J_{1/2-a}(n t_2) \gamma_n^{-1}\Bigr),\hspace{10mm}t_1<0,t_2>0,
\end{array}
\label{exprClimit}
\ee
where $\gamma_n$ are numbers originating from the coefficients of the power law expansion of the Bessel functions.

Note that the expression (\ref{exprClimit}) has the same algebraic structure as the one derived for the center-of-mass motion in \cite{EvninNguyen}, except that $|t|^a$ and $|t|^{1-a}$ are replaced by $\sqrt{|t|}J_\alpha(|t|)$ and $\sqrt{|t|}J_{-\alpha}(|t|)$. Demanding that the $\epsilon\rightarrow 0$ limit should exist results in the condition
\be
Q_{21}(\lambda)=0.\label{condLambda}
\ee
It is exactly the same condition as the one for the existence of a singular limit of the center-of-mass motion (generically leading to a discrete spectrum for $\lambda$). Under the assumption of (\ref{condLambda}) we obtain in the singular limit
\begin{align}
\mathcal{C}(t_1,t_2)&=\sqrt{-t_1 t_2}\,\frac{Q_{22}(\lambda) J_{a-1/2}(-n t_1)J_{1/2-a}(n t_2)-Q_{11}(\lambda) J_{1/2-a}(-n t_1)J_{a-1/2}(n t_2) }{W[\sqrt{-t_1}J_{a-1/2}(-n t_1),\sqrt{-t_1}J_{1/2-a}(-n t_1)]},\nonumber\\
&\hspace{10mm}t_1<0,t_2>0.\label{exprCtot}
\end{align}
The matching conditions across the singularity can now be derived rigorously by constructing two independent solutions to (\ref{de}). Note that all the information necessary for such construction is encoded (cf. (\ref{TifoC})) in the ``compression factor'' given by (\ref{exprCtot}). A convenient shortcut for this procedure is to recall the representation (\ref{repCfh}) of $\mathcal{C}(t_1,t_2)$ in terms of two arbitrary independent solutions $f(t)$ and $h(t)$, and to read off the corresponding singular limit of the two solutions directly from (\ref{exprCtot}). Writing $Q_{11}(\lambda)=q$ and $Q_{22}(\lambda)=-1/q$, we obtain as a basis of solutions,
\begin{align}
&Y_{1}(t)=\sqrt{-t}J_{a-1/2}(-n t),\hspace{5mm}Y_{2}(t)=\sqrt{-t}J_{1/2-a}(-n t),\hspace{5mm}t<0,\nonumber\\
&Y_{1}(t)=q\sqrt{t}J_{a-1/2}(n t),\hspace{5mm}Y_{2}(t)=-\frac{\sqrt{t}}{q}J_{1/2-a}(n t),\hspace{5mm}t>0.\label{basissol}
\end{align}

%%%%%%%%%%%%%%%%%%%%%%%%%%%%%%%%%%%%%%%%%

\section{The singular limit for the entire string}
\label{total}

As we have seen in the previous section, for $k\lambda >0$, 
consistent propagation of the string
center-of-mass across the singularity guarantees that all excited string modes
also propagate in a consistent fashion. This is not sufficient, however,
to define a consistent evolution for the whole string, since even small
excitations of higher string modes can sum up to yield an infinite total
energy \cite{HorowitzSteif}. As we shall see below, the condition of finite total string energy
(after the singularity crossing) turns out to be very restrictive.

The total string excitation energy can be conveniently expressed in terms
of the Bogoliubov coefficients for the higher string modes.
To compute the latter, we shall form two different bases of solutions from (\ref{basissol}) corresponding to purely positive and negative frequencies
at large negative and large positive times. More specifically, using the asymptotic expansion for the Bessel functions
\be
J_{\pm\alpha}(x)\sim\sqrt{\frac{2}{\pi x}}\mathrm{cos}\left(x \mp \alpha \frac{\pi}{2}-\frac{\pi}{4}\right),\hspace{10mm}x\rightarrow\infty,
\ee
we construct
\be
\begin{bmatrix}\phi_1^-\\ \phi_2^-\end{bmatrix}=\frac{i}{\mathrm{sin}(\alpha\pi)}\begin{bmatrix}-\mathrm{exp}(i\alpha\pi/2-i\pi/4)&\mathrm{exp}(-i\alpha\pi/2-i\pi/4)\\\mathrm{exp}(-i\alpha\pi/2+i\pi/4)&-\mathrm{exp}(i\alpha\pi/2+i\pi/4)\end{bmatrix}\begin{bmatrix}Y_1(t)\\Y_2(t)\end{bmatrix},\label{modefLeft}
\ee
such that,
\be
\phi_1^-(t)\sim \sqrt{\frac{2}{\pi n}} \mathrm{exp}(i n t),\hspace{10mm} \phi_2^-(t)\sim\sqrt{\frac{2}{\pi n}}\mathrm{exp}(-i n t),\hspace{10mm}t\rightarrow-\infty. 
\ee
Analogously, we introduce
\be
\begin{bmatrix}\phi_1^+\\ \phi_2^+\end{bmatrix}=\frac{i}{q\,\mathrm{sin}(\alpha\pi)}\begin{bmatrix}\mathrm{exp}(-i\alpha\pi/2+i\pi/4)&q^2\mathrm{exp}(i\alpha\pi/2+i\pi/4)\\-\mathrm{exp}(i\alpha\pi/2-i\pi/4)&-q^2\mathrm{exp}(-i\alpha\pi/2-i\pi/4)\end{bmatrix}\begin{bmatrix}Y_1(t)\\Y_2(t)\end{bmatrix},\label{modefRight}
\ee
such that
\be
\phi_1^+(t)\sim \sqrt{\frac{2}{\pi n}} \mathrm{exp}(i n t),\hspace{10mm} \phi_2^+(t)\sim\sqrt{\frac{2}{\pi n}}\mathrm{exp}(-i n t),\hspace{10mm}t\rightarrow+\infty. 
\ee
The two bases are related by a matrix made of Bogoliubov coefficients $\alpha_n$ and $\beta_n$:
\be
\begin{bmatrix}\phi_1^+\\\phi_2^+\end{bmatrix}=\begin{bmatrix}\alpha_n&\beta_n\\ \beta_n^*&\alpha_n^*\end{bmatrix}\begin{bmatrix}\phi_1^-\\\phi_2^-\end{bmatrix}
\ee
For the Bogoliubov coefficients, we obtain the following expressions, independent of $n$:
\begin{align}
\alpha_n&=-\frac{1+q^2}{2 q \,\mathrm{sin}(\alpha\pi)},\\
\beta_n&= i\frac{\mathrm{exp}(-i\pi \alpha)+q^2\mathrm{exp}(i\pi \alpha)}{2 q \,\mathrm{sin}(\alpha\pi)}.
\end{align}
Here, $\alpha=\sqrt{1-4k\lambda}/2$. The total mass of the string after crossing the singularity is given by \cite{HorowitzSteif}
\be
M=\sum_n n |\beta_n|^2.
\label{Msum}
\ee
Since the $\beta_n$ are $n$-independent, $M$ can only be finite\footnote{In general, one needs the uniformity of the $\epsilon\to 0$ limit of $\beta_n$ with respect to $n$ in order to analyze infinite sums as in (\ref{Msum}). As remarked at the end of section \ref{regII}, our considerations allow to draw immediate conclusions on the existence of the limit,
but not on its uniformity. However, since $M$ is a sum of positive numbers,
it is obvious that it will diverge when the $\beta_n$ approach an $n$-independent
non-zero value (in the $\epsilon\to 0$ limit), irrespectively of whether this approach
is uniform in $n$. For that reason, no further considerations are needed to draw our conclusions.} if $\beta_n=0$ for all $n$. For $k\lambda>0$, this cannot be achieved, since $0<\alpha<1/2$ and $q$ is real.
(For $k=0$, which is the case of the ``lightlike reflector plane'' of \cite{CDE},
all $\beta_n$ will vanish if $q^2=1$, which is satisfied automatically for any reflection-symmetric $\Omega(\lambda)$.)

%%%%%%%%%%%%%%%%%%%%%%%%%%%%%%%%%%%%%%%%%

\section{Discussion}
\label{discussion}

Before we recapitulate our main results, it shall be appropriate to make
two observations. 

First, one can ask what kind of cosmological singularities gives rise,
when the Penrose limit is taken,
to the plane wave singularities we have been considering.
According to \cite{plimit}, if one starts with isotropic homogeneous
cosmology of the type
\be
ds^2=-dt^2+t^{2h} dx^i dx^i,
\ee
and performs a Penrose limit, one obtains a plane wave of the form (\ref{genppwave}-\ref{scaleinv}) with
\be
k\lambda=\frac{h}{(1+h)^2}.
\ee
Thus, positive values of $k\lambda$ correspond to positive $h$, i.e., Friedmann-like Big Bang singularities, and negative values of $k\lambda$ correspond to negative $h$,
i.e., an infinite-expansion rather than an infinite-contraction singularity
(``Big Rip'').

Second, the dilaton field (discussed in more detail in appendix B) in the
backgrounds of the type (\ref{genppwave}-\ref{scaleinv}) takes the form \cite{PRT}
\be
\phi=\phi_0+cX^++\frac{dk\lambda}2\ln X^+
\label{dltn}
\ee
If $k\lambda$ is negative, this expression blows up near $X^+$ (and so does the string coupling) posing a serious threat
to the validity of perturbative string theory, and of free string propagation
as zeroth order approximation thereto. 

For this reason of limited validity of the free string approximation
when $k\lambda<0$, we have paid relatively little attention to this case.
What we could see is that, generically, it is hard to make excited string modes
propagate consistently across the singularity (though it may still be
possible to arrange such propagation by means of a judicious choice of the
resolved profile $\Omega(\eta)$ of the plane wave). The issue, however, cannot
be competently addressed within perturbative string theory on account
of string coupling blow-up. Our considerations can
be seen as a motivation to study these backgrounds
in the context of non-perturbative matrix theory descriptions of quantum gravity
(the Matrix Big Bang case of \cite{MBB} corresponds to the 11-dimensional
analog of the plane waves we have been considering compactified on a light-like circle with $k$ taken to $-\infty$ \cite{ppmatrix}.). Some steps in this direction have been
taken in \cite{ppmatrix}. (Alternatively, one could try to construct plane
wave backgrounds of the type (\ref{genppwave}-\ref{scaleinv}) where the curvature
of the metric is compensated by non-zero $p$-forms, rather than the dilaton,
thus avoiding the dilaton blow-up problem.)

For the case of positive $k\lambda$, i.e., those plane waves
that arise as Penrose limits of Friedmann-like cosmologies, it turns out
that individual excited string modes propagate consistently across the
singularity, whenever the center-of-mass of the string does. In those cases,
the dilaton (\ref{dltn}) is actually very large and {\it negative} near the singularity,
and one can expect that free strings are a good approximation
as far as propagation across the singularity is concerned (the string coupling
is small in the near-singular region). However, for free strings, we
find it impossible to maintain a finite total string energy after
the singularity crossing, provided
that the (scale-invariant) singularity is resolved in a way
that does not introduce new dimensionful parameters. The only way
out appears to be to allow hidden scales buried at the singular locus\footnote{If arbitrary resolutions, more general than (\ref{scaleinvres}), are allowed, for a given string mode, one should be able to reproduce (virtually) any matching conditions. This can be seen by assuming a particular form of solutions to the harmonic oscillator equation describing string propagation, and then reconstructing the plane wave profile necessary to produce this assumed motion. However, it is non-trivial to fit matching conditions for the entire tower of string modes in a particular geometrical resolution. For example, it is not obvious whether the matching conditions postulated in \cite{PRT} should have any geometrical interpretation at all.}
(even though the space-time away from the singularity is scale-invariant).
To contemplate the possible physical origins of such dimensionful scales
is an interesting pursuit, outside the scope of the present publication.

Another relevant consideration would be the propagation of strings across
plane wave singularities stronger than $1/(X^+)^2$. Unfortunately, at present,
little can be said about this case, even for the center-of-mass motion.

%%%%%%%%%%%%%%%%%%%%%%%%%%%%%%%%%%%%%%%%%

\section{Acknowledgments}

We would like to thank Matthias Blau for stimulating discussions, and Tim Nguyen for collaboration on closely related subjects. This research has been
supported in part by the Belgian Federal Science Policy Office through the Interuniversity Attraction Pole IAP VI/11, by the European
Commission FP6 RTN programme MRTN-CT-2004-005104 and by FWO-Vlaanderen through project G.0428.06.

%%%%%%%%%%%%%%%%%%%%%%%%%%%%%%%%%%%%%%%%%

\appendix

\section{Divergences for the case of the inverted harmonic oscillator}

As remarked in section \ref{eff_mat_cond}, for the case of $k\lambda <0$ 
(inverted harmonic oscillator), divergences may arise
in the evolution of excited string modes. These divergences may be
seen via a blunt application of (\ref{Cexpr0}), but it will
be more instructive to make their algebraic structure more explicit.

To this end, we shall derive a slightly different representation
for the total ``compression factor'' in place of (\ref{Cexpr0}).
One can start by rewriting (\ref{repCfh}) as
\be
\mathcal{C}(t_1,t_2)=\frac{1}{W[f,h]}\begin{pmatrix}f(t_1)&h(t_1)\end{pmatrix}\begin{pmatrix}0&1\\-1&0\end{pmatrix}\begin{pmatrix}f(t_2)\\h(t_2)\end{pmatrix}
\label{Cmatr}
\ee
For any two sets of solutions $\{f,h\}$ and $\{F,H\}$, the following relation holds:
\be
\begin{pmatrix}f(t)\\h(t)\end{pmatrix}=\frac{1}{W[F,H]}\begin{pmatrix}W[f,H]&-W[f,F]\\W[h,H]&-W[h,F]\end{pmatrix}\begin{pmatrix}F(t)\\H(t)\end{pmatrix}.
\label{solrel}
\ee
One can then take four sets of solutions: one approximated by $\{\sqrt{-t}J_\alpha(-nt),\sqrt{-t}J_{-\alpha}(-nt)\}$ in region I, two approximated
by $\{Y_{1-}(t/\epsilon),Y_{2-}(t/\epsilon)\}$ and $\{Y_{1+}(t/\epsilon),Y_{2+}(t/\epsilon)\}$ in region II, and one approximated by $\{\sqrt{t}J_\alpha(nt),\sqrt{t}J_{-\alpha}(nt)\}$ in region III. One can then start with (\ref{Cmatr}) written with the first of these four sets of solutions.
In this representation, the functions featured in (\ref{Cmatr}) are easily evaluated
at $t_1<0$, but not at $t_2>0$. One
then
consecutively applies (\ref{solrel}), (\ref{Q}) and (\ref{solrel}) again to insert
the remaining three sets of solutions, with the Wronskians in (\ref{solrel})
being evaluated at the boundaries of sub-regions. In the resulting expression,
all the four sets of solutions occur only with the values of the arguments
for which we have convenient approximations to these solutions, and the total compression
factor can be evaluated. As a matter of fact, this is simply another way to write (\ref{Cexpr0}).

The divergent contributions to the total ``compression factor'' can be identified with
particular Wronskians emerging from (\ref{solrel}), when one constructs the total ``compression factor'' with the procedure outlined in the previous paragraph.
For example, at the boundary of regions I and II, the following Wronskian occurs:
\be
W[\sqrt{-t}J_{-\alpha}(-n t),Y_{2-}(t/\epsilon)]\Big|_{t=-t_\epsilon}
\ee
The leading terms of both functions featured in the Wronskian are proportional to $|t|^{1-a}$, and therefore cancel by virtue of antisymmetry of the Wronskian.
However, the sub-leading contributions have a different functional form and do not have
to cancel. For example, for $a>3/2$, one may consider the contribution from
the first sub-leading power-law correction to the Bessel function, and
the leading term in $Y_{2-}$. This term is proportional to
\be
W[|t|^{3-a},|t|^{1-a}]\sim t^{3-2a},
\ee
and furthermore it is not accompanied by any powers of $\epsilon$ in the total expression for $\mathcal{C}(t_1,t_2)$. For that reason, evaluating this term at
$t=-t_\epsilon$ and taking the $\epsilon\to 0$ limit will produce a divergence.

\section{Background consistency and the singular limit for the dilaton\label{bgcc}}

As we have seen in the course of main exposition, consistent free string
propagation turns out to impose extremely stringent constraints on the
treatment of scale-invariant dilatonic plane wave backgrounds. For that
reason, it was not crucial for our picture to explore further conditions
arising from supergravity equations of motion imposed on the background.
However, for methodological completeness, we shall present considerations
for the singular limit of the dilaton field, and examine how this condition
combines with propagation of individual string modes. These derivation
will not have much bearing on the outcome of the analysis in the main text,
but they may be useful for pursuing various modifications of our present set-up.

If a time-dependent dilaton is used to support the curvature of the metric (\ref{genppwave}-\ref{scaleinv}) in the context of string theory, the condition for conformal invariance of the world-sheet theory is given by \cite{PRT}
\be
R_{\mu\nu}=-2 D_\mu D_\nu \phi.
\label{confinvar}
\ee
We shall impose this equation for all $X^+$ in the resolved plane wave profile,
and then examine the singular limit of the solutions for the dilaton. This is in contrast to the approach in \cite{PRT}, where the background consistency conditions at the singular locus were not discussed. The condition for conformal invariance (\ref{confinvar}) leads to the equation
\be
\ddot\phi(t)=-\frac{\lambda d}{2\epsilon^2}\Omega(t/\epsilon)
\label{dilEq}
\ee
for the dilaton where $d$ is the number of transverse dimensions $X^i$. We want to consider the limit $\epsilon\rightarrow 0$ of the solution $\phi$ to this equation. In order for this limit to exist, the regularization $\Omega$ will have to fulfill extra conditions. Since, in the singular limit, the space-time is regular away from $X^+=0$, we can construct a solution $\phi(t)$ to the left of the singularity and another solution $\phi(t)$ to the right. The requirements for the singular limit of $\phi$ to exist then reduce to demanding that the jumps in $\phi(t)$ and in its first derivative $\dot\phi(t)$ are finite:
\begin{align}
\Delta \phi &=\int_{t_1}^{t_2} \dot\phi(t) dt = \Bigl[ t \dot\phi(t) \Bigr]_{t_1}^{t_2} -\int_{t_1}^{t_2} t \ddot\phi(t) dt\\ 
& =\Bigl[ t \dot\phi(t) \Bigr]_{t_1}^{t_2} +\frac{\lambda d}{2}  \int_{t_1/\epsilon}^{t_2/\epsilon} \eta \Omega(\eta) d\eta\label{Dphi}
\end{align}
and
\be
\Delta\dot\phi= -\int_{t_1}^{t_2} \frac{\lambda d}{2\epsilon^2}\Omega(t/\epsilon) dt=-\frac{\lambda d}{2\epsilon}\int_{t_1/\epsilon}^{t_2/\epsilon} \Omega(\eta) d\eta
\ee
Thus, $\Delta\dot\phi$ can only be finite if
\be
\int_{-\infty}^{+\infty} \Omega(\eta) d\eta=0.\label{condB}
\ee
If that is the case, the first term in (\ref{Dphi}) is automatically finite,
and we are left to demand finiteness of the second term
\be
\lim\limits_{\epsilon\to 0}\int_{t_1/\epsilon}^{t_2/\epsilon} \eta \Omega(\eta) d\eta <\infty.
\ee
If $\Omega$ is even and satisfies (\ref{asrefprof}), this second condition is automatically satisfied.

%%%%%%%%%%%%%%%%%%%%%%%%%%%%%%%%%%%%%%%%%

\subsection{An explicit example}

We would now like to show that it is possible to combine the finite dilaton condition (\ref{condB}) with consistent propagation of individual string modes.
Given the considerations in the main text, this translates into finding
$\Omega(\eta)$ such that (\ref{condB}) is satisfied and, in addition,
\be
\frac{\partial^2}{\partial \eta^2} Y(\eta) + \lambda\Omega(\eta) Y(\eta) = 0\label{startEquation}
\ee
has a solution approaching $Y(\eta)\propto\eta^{1-a}$ for $\eta\rightarrow\pm\infty$.
We shall apply inverse reconstruction to $\Omega(\eta)$, assuming some shape
of this solution and adjusting it so as to satisfy
\be
\int_{-\infty}^{+\infty} \frac{Y''(\eta)}{Y(\eta)}\,d\eta=0. \label{Ycond}
\ee
This ``inverse reconstruction'' technique is generally useful for contemplating
qualitative properties of various plane wave profiles in relation to the singular limit.

\subsubsection{No-go theorem for $Y(\eta)$ without zero crossings}

In constructing an appropriate $Y(\eta)$, it is important to decide whether it should have zeros. If $Y$ has no zeros, $Y'/Y$ is regular everywhere, and we can rewrite (\ref{Ycond}) as:
\be
\int_{-\infty}^{+\infty} \frac{Y''(\eta)}{Y(\eta)}\,d\eta=\left[\frac{Y'(\eta)}{Y(\eta)}\right]^{+\infty}_{-\infty}+\int_{-\infty}^{+\infty} \frac{Y'^2(\eta)}{Y^2(\eta)}\,d\eta.
\ee
We now use $Y(\eta)\propto\eta^{1-a}$ for $\eta\rightarrow\pm\infty$, yielding
\be
\int_{-\infty}^{+\infty} \frac{Y''(\eta)}{Y(\eta)}\,d\eta=\int_{-\infty}^{+\infty} \frac{Y'^2(\eta)}{Y^2(\eta)}\,d\eta>0.
\ee
Therefore, if $Y$ has no zeros, it is impossible to construct an $\Omega(\eta)$ that integrates to zero. One must permit zeros (say $Y(\eta_i)=0$), and it is necessary to have bending points ($Y''(\eta_i)=0$) at the same locations due to (\ref{startEquation}). We will aim at constructing a symmetric $\Omega$, assuming that $Y$ is symmetric and restricting our analysis to $\eta>0$, and we will look for $Y$ that has
only one zero for $\eta>0$.

\subsubsection{Piece-wise construction of solution}

We will now prove that it is possible to construct an $\Omega$ that integrates to zero for a $Y$ that has one zero-crossing. $\Omega$ can be made arbitrarily smooth but for the simplicity of the proof we will allow $\Omega$ to have discontinuities. The main idea is to split the contributions to the integral 
\be\int_0^\infty \Omega(\eta)\mathrm{d}\eta,\label{totalOmega} \ee
into two parts, separated by $\eta=\eta_M$. The part 
\be \int_{\eta_M}^\infty \Omega(\eta)\mathrm{d}\eta,
\ee
will be chosen to be always positive. Then we prove that the contribution
\be \int_0^{\eta_M} \Omega(\eta)\mathrm{d}\eta\label{firstOmega},
\ee
can be made equal to any negative number while keeping the $\eta>\eta_M$ region intact. Therefore the total sum (\ref{totalOmega}) can always be taken zero by adjusting the $\eta<\eta_M$ contribution.

We rewrite equation (\ref{startEquation}) as
\be
\Omega=-\frac{1}{\lambda}\frac{Y''}{Y},
\ee
and we take a piecewise $Y(\eta)$ (with a continuous first derivative),
\be
Y(\eta)=\begin{cases} Y_1(\eta) \hspace{10mm}-\eta_M<\eta<\eta_M\\Y_2(\eta) \hspace{10mm}|\eta|>\eta_M.\end{cases}
\ee
The function $Y_2$ is fixed throughout our considerations, and we demand that it asymptotes to the subdominant solution for large $\eta$: $Y_2 \rightarrow \eta^{1-a}$ with $2a=1+\sqrt{1-4\lambda}$. As mentioned above, because of the denominator $Y$ in $\Omega$ there needs to be a bending point for each crossing of the $\eta$-axis. $Y_2''/Y_2$ is {\it negative} everywhere at $\eta>\eta_M$. The splicing point $\eta_M$ is taken to be a minimum, and we demand that $Y_1(\eta_M)=Y_2(\eta_M)\equiv Y(\eta_M)$. We take the following ansatz:
\be
Y_1(\eta)=(C-Y(\eta_M))\left(\frac{\eta^4}{\eta_M^4}-2\frac{\eta^2}{\eta_M^2}\right)+C.
\ee
A pictorial representation of our assumed solution is given on Fig.~\ref{fgAppB}.
\begin{figure}
\centering
\epsfig{file=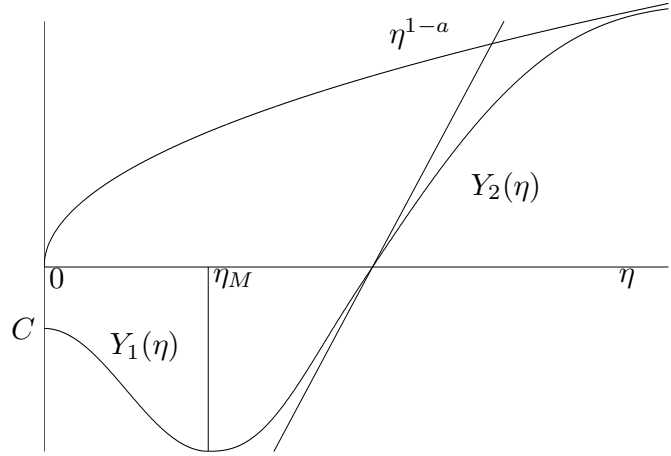, width=4in}
\caption{Piece-wise construction of $Y(\eta)$.}
\label{fgAppB}
\end{figure}
Due to the piecewise construction of $Y$ it is clear that $\int \Omega(\eta) d\eta$ consists of a separate $Y_1$ and $Y_2$ contribution. The contribution of $Y_2$ (i.e., -$\int_{\eta_M}^\infty Y_2''/Y_2 d\eta$) will always be positive. It remains to be proven that $Y_1$ can contribute an arbitrarily negative value for fixed $Y(\eta_M)$ and $\eta_M$. With $\eta_M>0$ and $\lambda>0$, this is equivalent to asking that
\be
\displaystyle \int_0^1 \frac{3 y^2-1}{y^4-2y^2+\frac{C}{C-Y(\eta_M)}}dy
\ee
can be set equal to an arbitrarily positive number. We know that $Y(\eta_M)\le C<0$, since $Y_1$ should not cross the $\eta$-axis and $\eta=\eta_M$ is a minimum. First, if $C=Y(\eta_M)$, the integral above is 0.
Then, for $C\rightarrow0^-$, with $\delta=-C/(C-Y(\eta_M))>0$, we find in the limit of $\delta \rightarrow 0$:
\be
\int_0^1 \frac{3 y^2-1}{y^4-2y^2-\delta}dy\sim \frac{\pi}{2\sqrt{2\delta}}.
\ee
For $C\to 0^-$ or $\delta\rightarrow0$ this becomes arbitrarily large and positive. As a consequence (\ref{firstOmega}) can be made equal to any negative number (between 0 and $-\infty$), and (\ref{condB}) can be satisfied
by appropriately adjusting $Y_1(\eta)$.

%%%%%%%%%%%%%%%%%%%%%%%%%%%%%%%%%%%%%%%%%%%%%%%%%%%%%%%%%%%%%%%%%%%%%%%%%%%%%%%%%%%%%%%%%%%%%%%%%%%%%%%%%%%%%%%%

%%%%%%%%%%%%%%%%%%%%%%%%%%%%%%%%%%%%%%%%%%%%%%%%%%%%%%%%%%%%%%%%%%%

\begin{thebibliography}{99}

\bibitem{HorowitzSteif} G.~T.~Horowitz and A.~R.~Steif, ``Space-Time Singularities in String Theory,'' Phys. Rev. Lett. {\bf 64} (1990) 260; G.~T.~Horowitz and A.~R.~Steif, ``Strings in Strong Gravitational
Fields,'' Phys. Rev. D {\bf 42} (1990) 1950.

\bibitem{PRT} G.~Papadopoulos, J.~G.~Russo and A.~A.~Tseytlin, ``Solvable model of strings
in a time-dependent plane-wave background,'' Class. Quant. Grav. {\bf 20} (2003) 969,
\texttt{arXiv:hep-th/0211289}.

\bibitem{AmatiKlimcik} D.~Amati and C.~Klim$\mathrm{\check{c}}$\'ik, ``Strings in a shock wave background and generation of curved geometry from flat-space string theory'', Phys. Lett. B {\bf 210} (1988) 94; D.~Amati and C.~Klim$\mathrm{\check{c}}$\'ik, ``Nonperturbative computation of the Weyl anomaly for a class of nontrivial backgrounds'', Phys. Lett. B {\bf 219} (1988) 443. 

\bibitem{MBB} B.~Craps, S.~Sethi and E.~P.~Verlinde,
  ``A matrix big bang,''
  JHEP {\bf 0510}, 005 (2005)
  \texttt{arXiv:hep-th/0506180},  B.~Craps, A.~Rajaraman and S.~Sethi,
  ``Effective dynamics of the matrix big bang,''
  Phys.\ Rev.\  D {\bf 73} (2006) 106005,
  \texttt{arXiv:hep-th/0601062}.

\bibitem{ppmatrix}M.~Blau and M.~O'Loughlin,
  ``DLCQ and Plane Wave Matrix Big Bang Models,''
  JHEP {\bf 0809} (2008) 097,
  \texttt{arXiv:0806.3255 [hep-th]}.
  
\bibitem{BMN}D.~E.~Berenstein, J.~M.~Maldacena and H.~S.~Nastase,
  ``Strings in flat space and pp waves from N = 4 super Yang Mills,''
  JHEP {\bf 0204} (2002) 013,
  {\tt arXiv:hep-th/0202021}.

\bibitem{plimit} M.~Blau, M.~Borunda, M.~O'Loughlin and G.~Papadopoulos, ``Penrose limits and spacetime singularities,'' Class. Quant. Grav. {\bf21} (2004) L43, \texttt{arXiv:hep-th/0312029}; M.~Blau, M.~Borunda, M.~O'Loughlin and G.~Papadopoulos,``The universality of Penrose limits near space-time singularities,'' JHEP {\bf 0407} (2004) 068, \texttt{arXiv:hep-th/0403252}.

\bibitem{David}
  J.~R.~David,
  ``Plane waves with weak singularities,''
  JHEP {\bf 0311} (2003) 064,
  {\tt arXiv:hep-th/0303013}.

\bibitem{EvninNguyen}O.~Evnin and T.~Nguyen, ``On discrete features of the wave equation
in singular pp-wave backgrounds,'' JHEP {\bf 0809} (2008) 105,
\texttt{arXiv:0806.3057 [hep-th]}.

\bibitem{RT}J.~G.~Russo and A.~A.~Tseytlin,
  ``A class of exact pp-wave string models with interacting light-cone gauge
  actions,''
  JHEP {\bf 0209} (2002) 035,
  {\tt arXiv:hep-th/0208114}.

\bibitem{CDE} B.~Craps, F.~De Roo and O.~Evnin, ``Quantum evolution across singularities: the case of geometrical resolutions,'' JHEP {\bf 04} (2008) 036, \texttt{arXiv:0801.4536 [hep-th]}.

\bibitem{gronwall}Anton Zettl, {\it Sturm-Liouville theory}, American Math. Society (2005), Chapter 1.

\end{thebibliography}
\end{document}